\documentclass[twocolumn,prb,showpacs,multicol,amsmath,amssymb]{revtex4}
\usepackage[dvips]{graphicx}
\usepackage{graphicx}% Include figure files
\usepackage{dcolumn}% Align table columns on decimal point
\usepackage{bm}% bold math
\usepackage{graphics}
\usepackage{epsfig,color}

\newcommand{\be}{\begin{equation}}
\newcommand{\ee}{\end{equation}}
\newcommand{\bea}{\begin{eqnarray}}
\newcommand{\eea}{\end{eqnarray}}

\begin{document}

\title{1D Frustrated Ferromagnetic Model with Added Dzyaloshinskii-Moriya Interaction}

\author{Javad Vahedi$^{1}$, Saeed Mahdavifar$^{2}$}
\affiliation{ $^{1}$Physics Research Center, Science Research Branch, Islamic Azad University, 19585-466, Tehran, Iran}
\address{$^{2}$ Department of Physics, University of Guilan, 41335-1914,Rasht, Iran}
\date{\today}

\begin{abstract}
 The one-dimensional (1D) isotropic frustrated ferromagnetic spin-1/2 model is considered. Classical and quantum effects of adding  a Dzyaloshinskii-Moriya (DM) interaction on the ground state of the system is studied using the analytical cluster method and numerical Lanczos technique. Cluster
 method  results, show that the classical ground state magnetic phase
 diagram consists of only one single phase: "chiral". The quantum corrections are determined by means of the Lanczos method and a rich quantum phase diagram including the gapless Luttinger liquid, the gapped chiral and dimer orders
 is obtained.  Moreover, next nearest neighbors will be entangled by increasing DM interaction and for open chains, end-spins  are entangled which shows the long distance entanglement (LDE) feature that can be controlled by DM interaction.
\end{abstract}


\maketitle

%%%%%%%%%%%%%%%%%%%%%%%%%%%%%%%%%%%%%%%%%%%%%%%%%%%%%%%%%%%%%%%%%%%%%%%%%%
%%%%%%%%%%%%%%%%%%%%%%%%%%     Section I      %%%%%%%%%%%%%%%%%%%%%%%%%%%%
%%%%%%%%%%%%%%%%%%%%%%%%%%%%%%%%%%%%%%%%%%%%%%%%%%%%%%%%%%%%%%%%%%%%%%%%%%

\section{Introduction}\label{sec1}

An important goal in study of quantum spin systems is the
search for novel phases emerging from competing interactions
between particles. Low dimensional quantum spin systems typically
exhibit strongly correlated effects which drive them toward new regimes
with no classical analog. Many properties of the systems in these
regimes or "quantum phases" can be understood if we explore their ground state and low-lying excitations.

Therefore a significant fraction of current research on such systems has focussed
on understanding of frustrated systems, which exhibits a variety of exotic quantum states\cite{new1}. In particular, the frustrated spin chains
are described by competing interactions between the nearest-neighbor ($J_{1}$) and
next-nearest-neighbor ($J_{2}$) interactions. The Hamiltonian of the
model is given by

\begin{equation}
H=\sum_{j=1}^{N}\left[J_{1}\vec{S}_{j}.\vec{S}_{j+1}+J_{2}\vec{S}_{j}.\vec{S}_{j+2}\right],
\label{J1-J2}
\end{equation}
%%%%%%%%%%%%%%%%%%%%%%%%%%%%%%%%%%
%%%%%%%%%%%%%%%%%%%%%%%%%%%%%%%%%%
where $\textbf{S}_{j}$ represents the $S=1/2$ operator at $j$-th
site of the chain. The model with both antiferromagnetic interactions $J_{1},J_{2}>0$  is
well studied \cite{Haldane82, Tonegawa87, Nomura92, Bursill95,
Majumdar69, White96, Bouzerar98, Kumar10-1, Kumar10-2}. Also the model (\ref{J1-J2}) with ferromagnetic and antiferromagnetic interactions
($J_{1}<0,J_{2}>0$) (frustrated ferromagnetic model) has been a subject of many studies \cite{Tonegawa89,
Chubukov91, Cabra00, Krivnov96}.
It is known that the
ground state is ferromagnetic for $\alpha=\frac{J_{2}}{|J_{1}|}<\frac{1}{4}$. At $\alpha_{c}=\frac{1}{4}$
the ferromagnetic state is degenerate with a singlet state. The
wave function of this singlet state is exactly known \cite{Hamada88, Dmitriev97}.
For $\alpha>\frac{1}{4}$, the ground state is an incommensurate singlet. It has been long
believed that at $\alpha>\frac{1}{4}$ the model is
gapless\cite{White96, Allen97} but the one-loop renormalization
group analysis indicates\cite{Cabra00, Nersesyan98-1} that the gap opens to a Lorentz symmetry breaking perturbation. However,
existence of the energy gap has not been yet verified
numerically\cite{Cabra00}. Using field theory
considerations it has been proposed\cite{Itoi01} that a very
tiny but finite gap exists which can be hardly observed by numerical techniques.

In vicinity of $\alpha=\frac{1}{4}$, the singlet ground state energy behaves as $E_{0}=(\alpha-\frac{1}{4})^{\beta}$, where $\beta$ is a critical exponent. Using variational
approaches\cite{Dmitriev07}, it has been shown that the quantum fluctuations definitely change
the classical critical exponent\cite{White96, Krivnov96} and yields $\beta=\frac{5}{3}$ which has been confirmed numerically\cite{Mahdavifar08-1}.

Beside a general interest in understanding {\em frustrations} and phase transitions in the
model systems described by the Hamiltonian in Eq(1), it helps people to understand intriguing
magnetic properties of a novel class of edge-sharing copper oxides, described by the
F-AF frustrated model\cite{Mizuno98, Hase04, Solodovnikov97}.
Recently some novel magnetic properties were discovered in a
variety of quasi-one dimensional materials that are known to
belong to the class of Dzyaloshinskii-Moriya (DM)
$(\overrightarrow{D}.(\vec{S}_{i}\times\vec{S}_{j}))$ magnet to
explain helical magnetic structures. The relevance of antisymmetric superexchange
interactions in spin Hamiltonians which describe quantum antiferromagnetic systems
was introduced phenomenologically by Dzyaloshinskii\cite{Dzyaloshinskii58}. Moriya
showed later, that such interactions arise naturally in the perturbation theory due
to the spin-orbit coupling in magnetic systems with low symmetry\cite{Moriya60}.

Some multiferroics cuprates,
such as $LiCuVO_{4}$\cite{l1,l2}, $LiCu_{2}O_{2}$\cite{l3} and $Cu_{2}GeO_{4}$,
are expected to be described by DM interaction.
This has stimulated extensive investigations
of various properties which are created by the DM
interaction. However, it is difficult to handle the DM interaction {\em analytically}
and interpret experimental data. A Numerical analysis then
helps us to understand experimental observation and even expand our knowledge about
many interesting quantum phenomena of low-dimensional quantum magnets. In the present work, we address a different problem in the subject
of the frustrated ferromagnetic spin-1/2 chains. We consider the
1D spin-1/2 frustrated ferromagnetic model with added DM
interaction and study the classical and quantum magnetic ground state phase diagram of the system.

The outline of the paper is as follows. In the next section the
classical cluster  method will be outlined and phases will be
obtained. In section III we present our numerical results of the
exact diagonalization calculations on the ground state properties
of the model. In section IV, the entanglement between different
spins will be investigated by calculating the concurrence
function. Finally we conclude and summarize
 our results in section V.

%%%%%%%%%%%%%%%%%%%%%%%%%%%%%%%%%%%%%%%%%%%%%%%%%%%%%%%%%%%%%%%%%%%%%%%%%%%%%%%%%%%%%%%%%%%%%%%%%%%%%%%%%%
%%%%%%%%%%%%%%%%%%%%%%%%%%%%%%%%%%%%%%%%%%%%%%%%%%%%%%%%%%%%%%%%%%%%%%%%%%
%%%%%%%%%%%%%%%%%%%%%%% BEGIN OF Section II %%%%%%%%%%%%%%%%%%%%%%%%%%
%%%%%%%%%%%%%%%%%%%%%%%%%%%%%%%%%%%%%%%%%%%%%%%%%%%%%%%%%%%%%%%%%%%%%%%%%%

\section{Classical phase diagram} \label{sec2}

The Hamiltonian of the 1D spin-1/2 frustrated ferromagnetic model
in presence of a uniform DM interaction is defined as

\begin{equation}
\emph{H}=\sum_{j=1}^{N} \left[
J_{1}\vec{S}_{j}.\vec{S}_{j+1}+J_{2}\vec{S}_{j}.\vec{S}_{j+2}+\vec{D}.(\vec{S}_{j}\times\vec{S}_{j+1})\right],
\label{J1-J2-DM}
\end{equation}
where $\overrightarrow{D}=D\hat{z}$. This Hamiltonian shows a
model with fully broken spin rotational symmetry. It is useful  to
begin with simple classical considerations that shed light on the
 possible ground state of the fully quantum mechanical problem.
 When $S$ is large, it is adequate to substitute
 $\vec{S}_{j}=S(\sin\theta_{j}\cos\phi_{j},\sin\theta_{j}\sin\phi_{j},\cos\theta_{j})$ in Eq.~(\ref{J1-J2-DM})
  and find solutions that satisfy
  $\partial\emph{H}/\partial\theta_{j}=0$ and $\partial\emph{H}/\partial\phi_{j}=0$
  for all $j$. But, here we use the rather unknown cluster method\cite{Lyons64,Kaplan09}
  of Lyons and Kaplan (LK method). Briefly recall that method with
  assuming periodic boundary condition. Then we can easily see that
   Eq.~(\ref{J1-J2-DM}) can be rewritten as:
%%%%%%%%%%%%%%%%%%%%%%%%%%%%
%%%%%%%%%%%%%%%%%%%%%%%%%%%%
\begin{equation}
\emph{H}_{c}=\sum_{j}h_{c}\Big(\vec{S}_{j-1},\vec{S}_{j},\vec{S}_{j+1}\Big),
\label{cluster-1}
\end{equation}
%%%%%%%%%%%%%%%%%%%%%%%%%%%%
%%%%%%%%%%%%%%%%%%%%%%%%%%%%
where the "cluster energy" involve three neighboring spins is
%%%%%%%%%%%%%%%%%%%%%%%%%%%%
%%%%%%%%%%%%%%%%%%%%%%%%%%%%
\begin{eqnarray}
h_{c}(\vec{S}_{1},\vec{S}_{2},\vec{S}_{3})&=&\frac{1}{2}\Big\{J_{1}(\vec{S}_{1}.\vec{S}_{2}+\vec{S}_{2}.\vec{S}_{3})\nonumber\\
&+&\vec{D}.(\vec{S}_{1}\times\vec{S}_{2}+\vec{S}_{2}\times\vec{S}_{3})\Big\}\nonumber \\
&+&J_{2}\vec{S}_{1}.\vec{S}_{3}.
\label{cluster-2}
\end{eqnarray}

%%%%%%%%%%%%%%%%%%%%%%%%%%%%
%%%%%%%%%%%%%%%%%%%%%%%%%%%%
It is clear that
%%%%%%%%%%%%%%%%%%%%%%%%%%%%
%%%%%%%%%%%%%%%%%%%%%%%%%%%%
\begin{equation}
\emph{H}_{c}\geq\sum_{j}min~h_{c}\Big(\vec{S}_{j-1},\vec{S}_{j},\vec{S}_{j+1}\Big),
\label{cluster-3}
\end{equation}
%%%%%%%%%%%%%%%%%%%%%%%%%%%%
%%%%%%%%%%%%%%%%%%%%%%%%%%%%
so easily one can find the minimum of $h_{c}$. If the
corresponding state propagates, i.e. if there is a state of the
whole system such that every set of three successive spins gives
the minimum $h_{c}$, then according to Eq.~(\ref{cluster-3}), this
state will be a ground state of the Hamiltonian. This is the main
idea of the LK cluster method as applied to the present problem.
This method has also been applied for models with higher dimensions
or with open boundary conditions\cite{Lyons64}.

%%%%%%%%%%%%%%%%%%%%%%%%%%%%%%%%%%%%%%%%%%%%%%%%%%%%%%%%%%%%%%%%%%%%%%%%%
%%%%%%%%%%%%%%%%%%%%%%% BEGIN OF THE FIGURE N 1 %%%%%%%%%%%%%%%%%%%%%%%%%%
%%%%%%%%%%%%%%%%%%  Figure-1  %%%%%%%%%%%%%%%%%%%%%%%%%%%%%%%%%
\begin{figure}[t]
\includegraphics[width=.85\columnwidth]{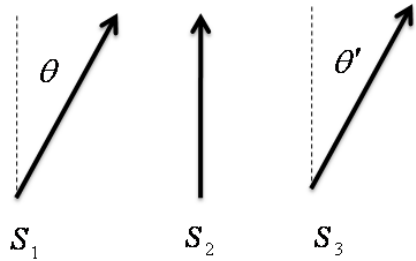}
\caption{(color online). The schematic picture of angles made by the spins in a
cluster of three spins.}\label{schematic}
\end{figure}
%%%%%%%%%%%%%%%%%%%%%%%%%%%%%%%%%%%%%%%%%%%%%%%%%%%%%%%%%%%%%%%

%%%%%%%%%%%%%%%%%%%%%%%%%%%%%%%%%%%%%%%%%%

Now let's minimize $h_{c}$. Because the DM vector is selected
along the $z$ axis, we consider only coplanar (spin dimensionality
d=2, i.e. XY spins) state. Without loss of generality label the
angles made by the end spins with the central spin $\theta$,
$\theta'$ (Fig.\ref{schematic}). So the cluster energy is became
%%%%%%%%%%%%%%%%%%%%%%%%%%%%
%%%%%%%%%%%%%%%%%%%%%%%%%%%%
\begin{eqnarray}
h_c(\theta,\theta')&=&\frac{|J_{1}|}{8}\Big\{-\cos\theta-\cos\theta'+\gamma(\sin\theta-\sin\theta') \nonumber \\
&+&2\alpha\cos(\theta-\theta')\Big\},
\label{cluster-4}
\end{eqnarray}
%%%%%%%%%%%%%%%%%%%%%%%%%%%%
%%%%%%%%%%%%%%%%%%%%%%%%%%%%
where $\gamma=\frac{D}{|J_{1}|}$. Differentiating gives the
conditions for stationarity
%%%%%%%%%%%%%%%%%%%%%%%%%%%%
%%%%%%%%%%%%%%%%%%%%%%%%%%%%
\begin{eqnarray}
\frac{|J_{1}|}{8}\Big\{\sin\theta+\gamma\cos\theta-2\alpha\sin(\theta-\theta')\Big\}&=&0\nonumber\\
\frac{|J_{1}|}{8}\Big\{\sin\theta'-\gamma\cos\theta'+2\alpha\sin(\theta-\theta')\Big \}&=&0.
\label{cluster-5}
\end{eqnarray}
%%%%%%%%%%%%%%%%%%%%%%%%%%%%
%%%%%%%%%%%%%%%%%%%%%%%%%%%%
Before going through the Eq.~(\ref{cluster-4}), we check two simplified cases as follow.
\\

%%%%%%%%%%%%%%%%%%%%%%%%%%%%%%%%%%%%%%%%%%%%%%%%%%%%%%%%%%%%%%%%%%%%%%%%%
%%%%%%%%%%%%%%%%%%%%%%% BEGIN OF THE FIGURE N 2 %%%%%%%%%%%%%%%%%%%%%%%%%%
%%%%%%%%%%%%%%%%%%%%%%%%%%%%%%%%%%%%%%%%%%%%%%%%%%%%%%%%%%%%%%%%%%%%%%%%%%
%%%%%%%%%%%%%%%%%%%%%%%%%%%%%%%%%%%%%%%%%

\begin{figure}[t]
\centerline{\psfig{file=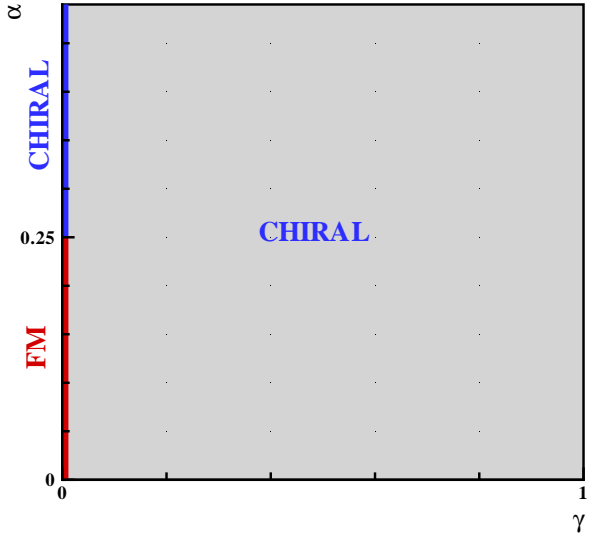,width=3.65in}}
\caption{(color online). The classical ground state magnetic phase diagram.}\label{classical}
\end{figure}
%%%%%%%%%%%%%%%%%%%%%%%%%%%%%%%%%%%%%%%%%%%%%%%%%%%%%%%%%%%%%%%

\textbf{First case}: We have put
$D=0$ and the frustrated ferromagnetic
 Heisenberg model recovered which well studied numerically and analytically.
 In this case, solutions are:
%%%%%%%%%%%%%%%%%%%%%%%%%%%%
%%%%%%%%%%%%%%%%%%%%%%%%%%%%
\begin{eqnarray}
(\theta,\theta')&=&(0,0),(\pi,\pi)(0,\pi)(\pi,0)~~~~~(Ising -type)\nonumber\\
(\theta,\theta')&=&(\theta_{0},-\theta_{0})~~~~~(chiral- type),~~~~~where\nonumber\\
\cos\theta_{0}&=&\frac{1}{4\alpha}.
\end{eqnarray}
%%%%%%%%%%%%%%%%%%%%%%%%%%%%
%%%%%%%%%%%%%%%%%%%%%%%%%%%%
The $(\pi, \pi)$ solution, which leads to the ordinary
antiferromagnetic state, is never lowest because we have assumed
$J_{1} < 0$. The $(0,0)$ solution obviously propagates as the
ferromagnetic state. The solutions $(\pi, 0)$, $(0, \pi)$,i.e.
$(\downarrow, \uparrow, \uparrow)$, $(\uparrow,
\uparrow,\downarrow )$ plus their degenerate reversed spin
counterparts can easily be seen to
 propagate in the up-up-down-down state\cite{Lyons64}. %The solution $(\theta_{0},-\theta_{0})$, degenerate with its

 We listed the energies for the various stationary solutions:
%%%%%%%%%%%%%%%%%%%%%%%%%%%%
%%%%%%%%%%%%%%%%%%%%%%%%%%%%
\begin{eqnarray}
h_{ferro}&=&h_{c}(0,0)=\frac{|J_{1}|}{8}(-2+2 \alpha),\nonumber\\
h_{uudd}&=&h_{c}(0,\pi)=\frac{|J_{1}|}{8}(-2 \alpha),\nonumber\\
h_{chiral}&=&h_{c}(\theta_{0},-\theta_{0})=\frac{|J_{1}|}{8}(-\frac{1}{4\alpha}-2
\alpha).
\end{eqnarray}
%%%%%%%%%%%%%%%%%%%%%%%%%%%%
%%%%%%%%%%%%%%%%%%%%%%%%%%%%
By equating these energies in pairs we found only one
critical point, $\alpha_{c}=0.25$. The ground state is in the
ferromagnetic phase for $\alpha<0.25$ and is in the chiral phase
for $\alpha>0.25$.
%\cite{Fisher80}.
\\

\textbf{Second case}: We have put
$J_{2}=0$ (or $\alpha=0$) which means we have just 1D isotropic ferromagnetic Heisenberg model
plus a uniform DM interaction. In this case there is only one solution:
%%%%%%%%%%%%%%%%%%%%%%%%%%
 \begin{eqnarray}
 (\theta,\theta')&=&(\theta_{0},-\theta_{0})~~(chiral~ type)~~~where \nonumber\\
 \tan\theta_{0}&=&-\gamma.
 \end{eqnarray}
 %%%%%%%%%%%%%%%%%%%%%%%%%
 Which means that immediately after turning on the DM interaction, the long range chiral order is created in
 the classical ground state phase diagram.  In principle, as soon as the DM interaction is applied, the ground state of the system undergoes a phase transition from the ferromagnetic phase into a chiral phase.

 It is better to emphasis that the induced effects of quantum fluctuations on the 1D antiferromagnetic Heisenberg model with DM interaction had been studied theoretically
 and experimentally\cite{Mahdavifar08-2, Soltani10, Garate10}. From quantum point of view, it has been found that the DM interaction
 induces the chiral phase which remains stable even in presence of a uniform magnetic field\cite{Mahdavifar08-2}.
\\

\textbf{Third case}: Now we
consider the 1D frustrated ferromagnetic model with DM interaction. The solutions for general equations (\ref{cluster-4}) are
%%%%%%%%%%%%%%%%%%%%%%%%%%%%%%%%%
\begin{eqnarray}
(\theta,\theta')&=&(0,0)~~~where~~~\gamma=0 \nonumber\\
(\theta,\theta')&=&(0,\pi)=(\pi,0)~~~where~~~\gamma=0\nonumber\\
(\theta,\theta')&=&(\theta_{0},-\theta_{0})~~~where\nonumber \\
4\alpha&=&\frac{1}{\cos\theta_{0}}+\frac{\gamma}{\sin\theta_{0}}.
\end{eqnarray}
%%%%%%%%%%%%%%%%%%%%%%%%%%%%%%%%%
 These results show, the ferromagnetic and up-up-down-down orders exist only in the
 absence of the DM interaction. As soon as the DM interaction
 increases from zero, the ground state of the system goes to the
 chiral phase, independent of the frustrated parameter $\alpha$ (Fig.\ref{classical}).
 Thus, using the cluster LK method, the classical ground state phase
 diagram of the 1D spin-1/2 frustrated ferromagnetic model with added
 uniform DM interaction consists of a single phase: "chiral".

%%%%%%%%%%%%%%%%%%%%%%%%%%%%%%%%%%%%%%%%%%%%%%%%%%%%%%%%%%%%%%%%%%%%%%%%%%
%%%%%%%%%%%%%%%%%%%%%%%%%%     Section IV     %%%%%%%%%%%%%%%%%%%%%%%%%%%%
%%%%%%%%%%%%%%%%%%%%%%%%%%%%%%%%%%%%%%%%%%%%%%%%%%%%%%%%%%%%%%%%%%%%%%%%%%
\section{Quantum phase diagram}\label{sec4}

In this section, to explore the nature of the spectrum and the
quantum phase transition, we used the Lanczos method to diagonalize
numerically chains with length up to $N=24$.

\begin{figure}[t]
\centerline{\psfig{file=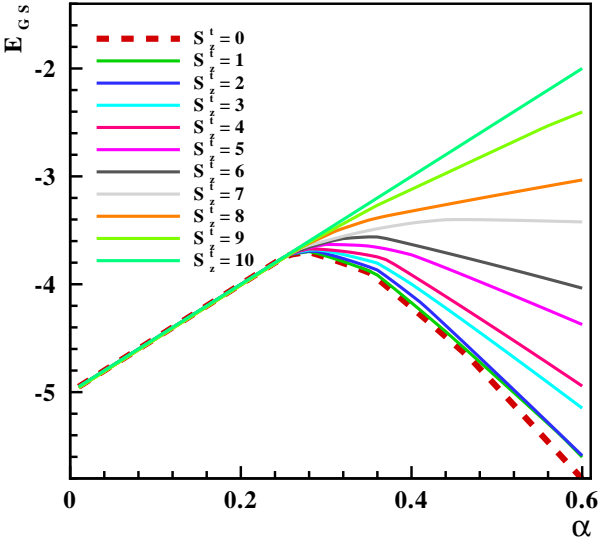,width=3.65in}}
\caption{(color online). Ground state energy of the
system in different subspaces versus the frustration parameter
$\alpha$ for chain with length $N=20$ and DM vector
$D=0.05$.}\label{energy-gap}
\end{figure}

First, in different subspaces  we have computed the ground state energy of
chains with $J_1=-1$ and different values of the DM interaction as
a function of the parameter $\alpha$. The setting of boundary
conditions and observation of energy with respect to change of
parameters is important for precise analysis of wave
state\cite{ATN05}, but because of our limitation and size effects,
we have just set the periodic boundary condition. In
Fig.~\ref{energy-gap}, we present results of these calculations
for the value of DM vector $D=0.05$ and chain size $N=20$. It can be seen that the
ground state energy for weak DM interaction is nearly degenerate
up to the first critical frustration $\alpha_{c_1}=0.24 \pm
0.01$, which value is obtained by extrapolation  technique. Also degeneracy in the mentioned region shows that in the absence of the
frustration, $\alpha=0$, the spectrum of the model is gapless. By
turning on the  frustration, spectrum remains gapless up to the
first critical value of the frustration, $\alpha_{c_1}=0.24 \pm
0.01$. As soon as the frustration increases from $\alpha_{c_{1}}$, the
ground state is non-degenerate and exists in the subspace with
total $S^{z}=0$. Fig.~\ref{energy-gap} is also good analysis,
especially for the proof of ferromagnetic sate.

%%%%%%%%%%%%%%%%%%%%%%%%%%%%%%%%%%%%%%%%%%%%%%%%%%%%%%%%%%%%%%%%%%%%%%%%%%
%%%%%%%%%%%%%%%%%%%%%%% BEGIN OF THE FIGURE N 1 %%%%%%%%%%%%%%%%%%%%%%%%%%
%%%%%%%%%%%%%%%%%%%%%%%%%%%%%%%%%%%%%%%%%%%%%%%%%%%%%%%%%%%%%%%%%%%%%%%%%%
%%%%%%%%%%%%%%%%%%%%%%%%%%%%%%%%%%%%%%%%%
%%%%%%%%%%%%%%%%%%%%%%%%%%%%%%%%%%%%%%%%

%%%%%%%%%%%%%%%%%%%%%%%%%%%%%%%%%%%%%%%%%%%%%%%%%%%%%%

From the viewpoint of the symmetry our model
is completely different than the models studied in the literature
\cite{Dmitriev08,Misner09,Furukawa10,Misner06,Vekua07,Kecke07,Hikihara08,Sudan09,Hikihara00,Hikihara01}.
The latter models have the $Z_{2}\times U(1)$ symmetry. In this case,
the absence of the chiral LRO
is natural, and the spontaneous breaking of the $Z_{2}$ chiral symmetry,
which resulted in the chiral LRO, is interesting and was indeed
the main topic in those studies.
On the other hand, in the present model, the $Z_{2}$ chiral symmetry is broken by
the DM interaction explicitly. Therefore, the appearance of the chiral
LRO should be natural; if the LRO is absent, it should be regarded
as a surprising result and there must be a very interesting mechanism
which recovers the symmetry behind the result.

%%%%%%%%%%%%%%%%%%%%%%%%%%%%%%%%%%%%%%%%%%%%%%%%%%%%%%%%%%%%%%%%%%%%%%%%%%
%%%%%%%%%%%%%%%%%%%%%%% BEGIN OF THE FIGURE N 2 %%%%%%%%%%%%%%%%%%%%%%%%%%
%%%%%%%%%%%%%%%%%%%%%%%%%%%%%%%%%%%%%%%%%%%%%%%%%%%%%%%%%%%%%%%%%%%%%%%%%%
%%%%%%%%%%%%%%%%%%%%%%%%%%%%%%%%%%%%%%%%%

\begin{figure}[t]
\centerline{\psfig{file=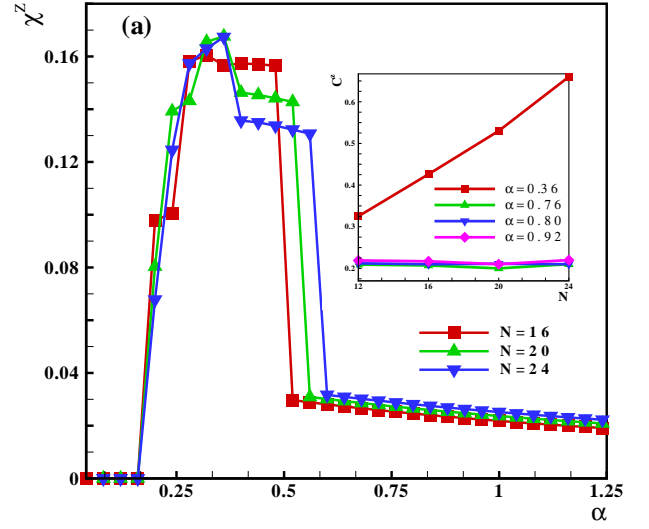,width=3.65in}}
\centerline{\psfig{file=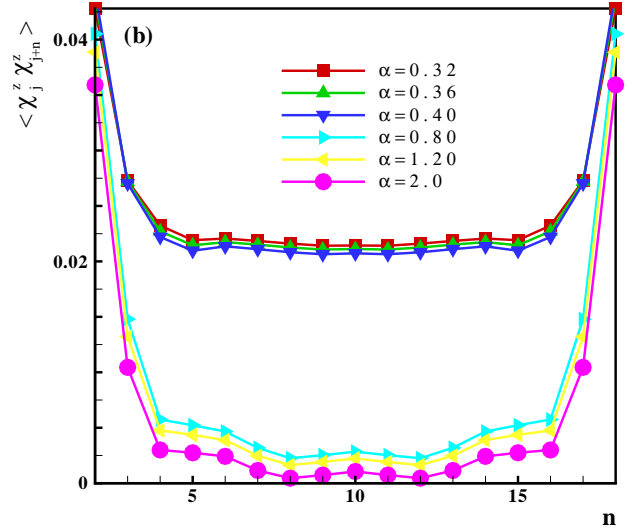,width=3.65in}}
\caption{(color online). (a) The chiral order parameter $\chi^{z}$ as
a function of the frustration parameter $\alpha$ for the value of
DM vector $D=0.05$, including different chain lengths $N=16, 20,
24$. In the inset the chiral correlation function is plotted as a
function of the chain length $N$ for the value of frustration
$\alpha=0.36$. The diverging behavior of the correlation function
shows that the chiral ordering in the intermediate region is true
long range. (b) The correlation $\langle \chi_{j}^{z}~\chi_{j+n}^{z}\rangle$ as
a function of $n$ for chain length $N=20$ and $D=0.05$.}\label{chiral-order}
\end{figure}
%%%%%%%%%%%%%%%%%%%%%%%%%%%%%%%%%%%%%%%%%%%%%%%%%%%%%%

%%%%%%%%%%%%%%%%%%%%%%%%%%%%%%%%%%%%%%%%%%
%***********************************************************
%%%%%%%%%%%%%%%%%%%%%%%%%%%%%%%%%%%%%%%%%%%%%%%%%%%%%%%%%%%%%%%%%%%%%%%%%%

In order to recognize the different quantum phases,
we have implemented the Lanczos algorithm on finite size chains to
calculate the lowest eigenstate. The first insight into the nature
of different phases can be obtained studying the chiral order
parameter. It has been shown that the DM interaction can create 
the chiral phase in the ground state phase diagram of the spin
systems\cite{Mahdavifar08-2, Soltani10}, which is characterized by 
nonzero value of the  chirality

%***********************************************************
\begin{eqnarray}
\chi^{z}=\frac{1}{N} \sum_{j}\langle \chi_{j}^{z}\rangle, \label{chirality}
\end{eqnarray}
%***********************************************************
where $\chi_{j}^{z}=({\bf S}_{j}\times {\bf S}_{j+1})^{z}$.
One should note that there are two different types of the  chiral
ordered phases, gapped and gapless\cite{Hikihara00,Kaburagi99}. In
Fig.~\ref{chiral-order}(a), we have presented calculated results on
the chiral order parameter as a function of the frustration
parameter $\alpha$ for a value of DM vector $D=0.05$, including
different chain lengths $N=16, 20, 24$. As is clearly seen from
this figure, there is no long-range chiral order along the $z$
axis in the regions $\alpha<\alpha_{c_1}$ and
$\alpha>\alpha_{c_2}$. However, in the intermediate region, $\alpha_{c_1}<\alpha<\alpha_{c_2}$, the ground state shows a
profound chiral order in the $z$ direction. Since the study of
correlation functions can give us deep insight into the
characteristics of the ground state, we define the chiral
correlation function as
%***********************************************************
\begin{eqnarray}
C^{z}=\sum_{n=1}^{N}\langle \chi_{j}^{z}~\chi_{j+n}^{z}\rangle.
\label{chiral-cor}
\end{eqnarray}
%***********************************************************
It is important that in a phase with true long-range order
the correlation functions should diverge as
$N\longrightarrow\infty$. To check the
existence of the chiral long-range
order in the thermodynamic limit $N \longrightarrow \infty$ of the
system, we have plotted in the inset
of Fig.~\ref{chiral-order}(a) the $N$ dependence of $C^{z}$ for different
values of frustration in the regions $\alpha>\alpha_{c_{1}}$. As is seen from this figure, only in
the intermediate region $\alpha_{c_1}<\alpha<\alpha_{c_2}$, there is a diverging behavior which shows
that the chiral order in the intermediate region is true
long-range order. On the other hand, the constant value in the region
$\alpha>\alpha_{c_{2}}$ shows that the $C^{z}/N$ takes zero value in the
thermodynamic limit $N\longrightarrow\infty$. For grater emphasis, we have
plotted in Fig.~\ref{chiral-order}(b), the correlation function
$\langle \chi_{j}^{z}~\chi_{j+n}^{z}\rangle$ as a function of $n$ for
a chain length $N=20$. Clearly is seen that the long distance correlation
of the chirality only exist in the region $\alpha_{c_1}<\alpha<\alpha_{c_2}$.
The mirror symmetry is the result of the periodic boundary condition.  Therefore,
by studying the ground state energy and the chiral order parameter, we
showed that in the intermediate region
$\alpha_{c_1}<\alpha<\alpha_{c_2}$, the 1D spin-1/2 frustrated
ferromagnetic model with added DM interaction is in the gapped
chiral order phase.

\begin{figure}[t]
\centerline{\psfig{file=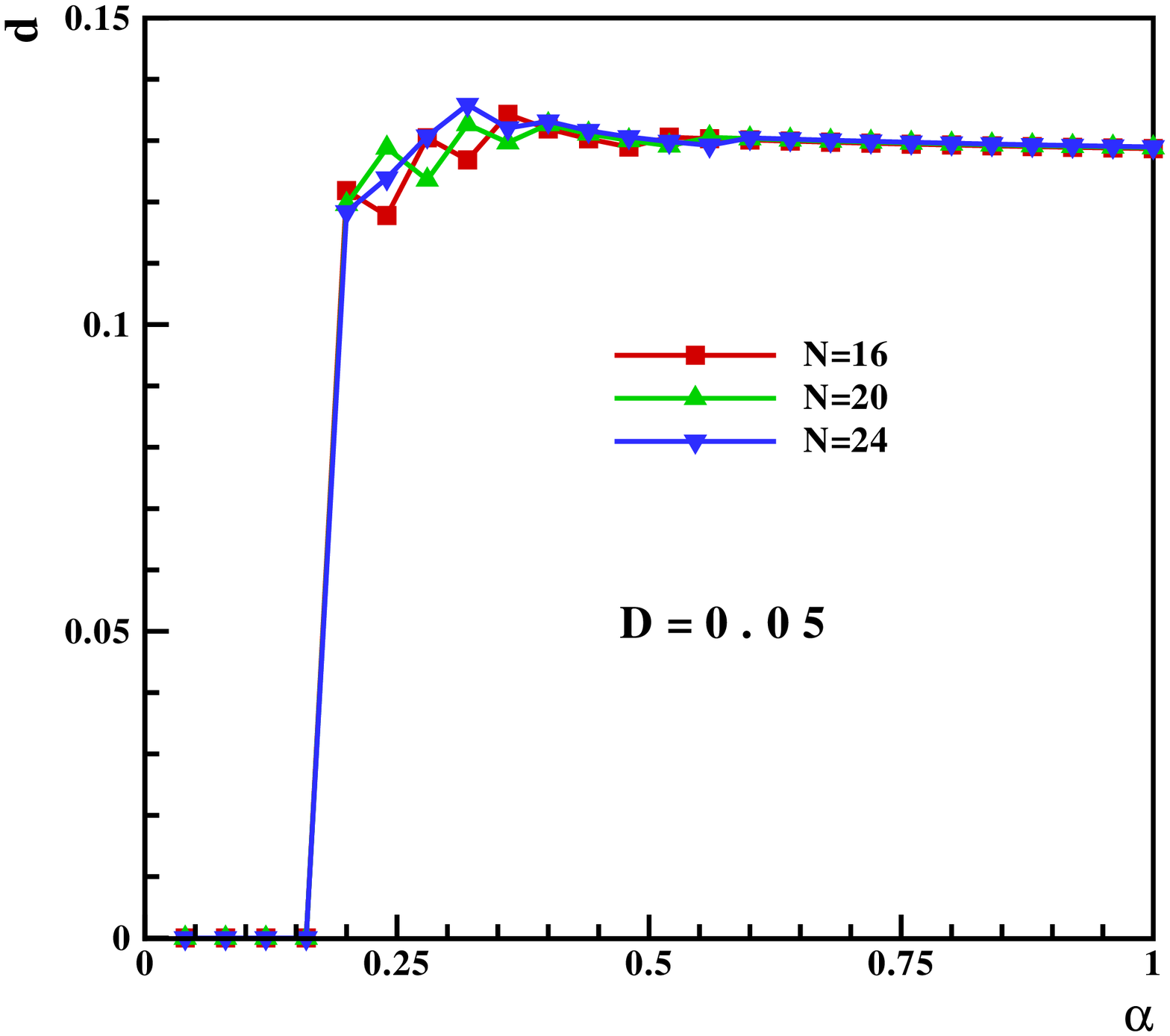,width=3.65in}}
\caption{(color online). The dimer order parameter, $d$, as a
function of the frustration parameter $\alpha$ for the value of DM
vector $D=0.05$, including different chain lengths $N=16, 20, 24$.
}\label{Dimer-order}
\end{figure}

In what follows we will present our numerical study about the presence of dimer
phase. In order to understand the nature of the dimer phase let us assume we have a
1D frustrated antiferromagnetic model, where the sign of $J_{1}$
will be changed through the $\pi$ rotations around the $z$ axis of the
spins on every second sites and the model will be transformed to the 1D frustrated
ferromagnetic model. From the fact that the ground state at the Majumdar-Ghosh
point ($J_{2}=J_{1}/2$) is given by the product of singlet dimmers ($\frac{1}{\surd
2}(|\uparrow\downarrow\rangle-|\downarrow\uparrow\rangle)$)\cite{Majumdar69}. Through
the above $\pi$-rotation transformation, the dimer unit for
$J_{2}=-J_{1}/2$ is replaced by the triplet state $\frac{1}{\surd
2}(|\uparrow\downarrow\rangle+|\downarrow\uparrow\rangle)$\cite{Chubukov91}.
To this reason, to find additional insight into the different
phases we are focused on existing of the dimer phase in our 1D frustrated ferromagnetic model. The order
parameter characterizing the dimer phase is given as\cite{White96}

%***********************************************************
\begin{eqnarray}
d = \frac{1}{N}\sum_{i}(-1)^{i}\langle S_{i}.S_{i+1}\rangle. \label{Dimer}
\end{eqnarray}
%***********************************************************
This order parameter shows alternating signs along the spin chain.
Our results for the DM vector value $D=0.05$ and different chain sizes
$N=16, 20, 24$ are presented in Fig.~\ref{Dimer-order}. As can
clearly be seen from this figure, in the Luttinger liquid region, $\alpha<\alpha_{c_1}$, there is not any long-range dimer order.
As soon as the frustration increases from the first critical
value, the dimer ordering increases from zero very rapidly. The
oscillations of the dimer order in the intermediate region at
finite $N$ are the result of level crossing between ground state
and excited states of the model. On the other hand, overlapping of
the numerical results in the region $\alpha>\alpha_{c_2}$, shows a
divergent behavior of the correlation function of the dimer order
parameter by increasing the size of chain $N$. This justifies that
the true long-range dimer order exists in the region
$\alpha>\alpha_{c_2}$ of the ground state phase diagram.

%%%%%%%%%%%%%%%%%%%%%%% BEGIN OF THE FIGURE N 3 %%%%%%%%%%%%%%%%%%%%%%%%%%
%%%%%%%%%%%%%%%%%%%%%%%%%%%%%%%%%%%%%%%%%%%%%%%%%%%%%%%%%%%%%%%%%%%%%%%%%%
%%%%%%%%%%%%%%%%%%%%%%%%%%%%%%%%%%%%%%%%%
%%%%%%%%%%%%%%%%%%%%%%%%%%%%%%%%%%%%%%%%%%%%%%%%%%%%%%

%%%%%%%%%%%%%%%%%%%%%% BEGIN OF THE FIGURE N 2 %%%%%%%%%%%%%%%%%%%%%%%%%%
%%%%%%%%%%%%%%%%%%%%%%%%%%%%%%%%%%%%%%%%%%%%%%%%%%%%%%%%%%%%%%%%%%%%%%%%%%
%%%%%%%%%%%%%%%%%%%%%%%%%%%%%%%%%%%%%%%%%
%%%%%%%%%%%%%%%%%%%%%%%%%%%%%%%%%%%%%%%%%
\begin{figure}[t]
\centerline{\psfig{file=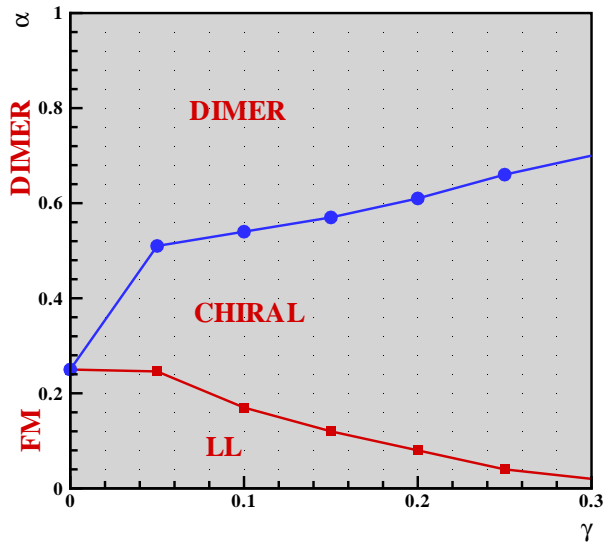,width=3.65in}}
\caption{(color online). The quantum ground state magnetic phase diagram.
}\label{quantum}
\end{figure}
%%%%%%%%%%%%%%%%%%%%%%%%%%%%%%%%%%%%%%%%%%%%%%%%%%%%%%

%%%%%%%%%%%%%%%%%%%%%%%%%%%%%%%%%%%%%%%%%%

%%%%%%%%%%%%%%%%%%%%%%%%%%%%%%%%%%%%%%%%%%%%%%%%%%%%%%%%%%%%%%%%%%%%%%%%%
%
We have to mention that for other values of the DM interaction
($D\leq0.3$), we did the same numerical experiment and found the
same quantum picture of the ground state magnetic phase diagram
contains of: (I.) gapless Luttinger liquid phase in the region
$\alpha<\alpha_{c_1}$ (II.) gapped chiral phase in the
intermediate region $\alpha_{c_1}<\alpha<\alpha_{c_2}$ and (III.)
gapped dimer phase for $\alpha>\alpha_{c_2}$ (Fig.\ref{quantum}).

%%%%%%%%%%%%%%%%%%%%%%%%%%%%%%%%%%%%%%%%%%%%%%%%%%%%%%%%%%%%%%%%%%%%%%%%%%
%%%%%%%%%%%%%%%%%%%%%%%%%%     Section V     %%%%%%%%%%%%%%%%%%%%%%%%%%%%
%%%%%%%%%%%%%%%%%%%%%%%%%%%%%%%%%%%%%%%%%%%%%%%%%%%%%%%%%%%%%%%%%%%%%%%%%%
\section{Concurrence}\label{sec V }

In this section we study the entanglement of formation as a
measure of the entanglement.
Entanglement stands some similarity  to classical
correlation, but it differs in some important respects, including the
fact that entangled objects can violate Bells inequality \cite{bell}.
May be one of the most characteristic differences is this: if
two similar quantum objects are completely entangled with
each other, then neither of them can be at all entangled with
any other object, but there is no such restriction on classical
correlations\cite{Wooters2001}.

We compute the entanglement between two sites which
is known as the concurrence\cite{Wooters98, Amico08}
%***********************************************************
\begin{eqnarray}
C_{lm}&=&2~max\left\{0, C_{lm}^{(1)}, C_{lm}^{(2)}\right\},
\label{concurrence-1}
\end{eqnarray}
%***********************************************************
where
%***********************************************************
\begin{eqnarray}
C_{lm}^{(1)}&=& \sqrt{(g_{lm}^{xx}-g_{lm}^{yy})^{2}+(g_{lm}^{xy}+g_{lm}^{yx})^{2}} \nonumber \\
&-&\sqrt{(\frac{1}{4}-g_{lm}^{zz})^{2}-(\frac{M_{l}^{z}-M_{m}^{z}}{2})^{2}}\nonumber \\
C_{lm}^{(2)}&=& \sqrt{(g_{lm}^{xx}+g_{lm}^{yy})^{2}+(g_{lm}^{xy}-g_{lm}^{yx})^{2}} \nonumber \\
&-&\sqrt{(\frac{1}{4}+g_{lm}^{zz})^{2}-(\frac{M_{l}^{z}+M_{m}^{z}}{2})^{2}},
\label{concurrence-2}
\end{eqnarray}
%***********************************************************
where $M^{z}$ is magnetization along $z$ axis and $g_{lm}^{\mu\nu}=\langle S_{l}^{\mu}
S_{m}^{\nu}\rangle$ is the correlation function between spins on
sites $l$ and $m$.

\begin{figure}[t]
\centerline{\psfig{file=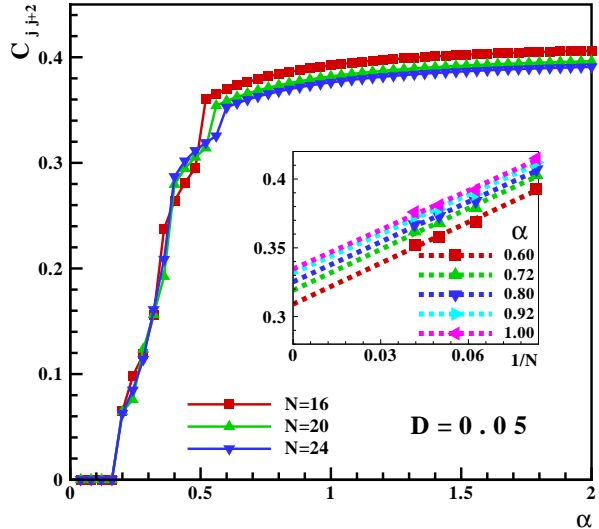,width=3.65in}}
\caption{(color online). The entanglement between next nearest
neighbors $C_{j~j+2}$ as a function of the frustration parameter
$\alpha$ for the value of DM vector $D=0.05$, including different
chain lengths $N=16, 20, 24$. The inset shows the concurrence as a
function of the inverse length $1/N$ for different values of the
frustration $\alpha=0.60, 0.72, 0.80, 0.92, 1.00$. Finite linear
extrapolated results show that in the thermodynamic limit
$N\longrightarrow \infty$, the next nearest neighbors are
entangled in the dimer phase.}\label{Concurrence}
\end{figure}

The numerical Lanczos results on the concurrence are shown in Fig.~\ref{Concurrence}. From our numerical results we
found that the concurrence of two spins that are nearest neighbor
is equal to zero in all regions of the quantum ground state phase diagram.
In this figure the concurrence of one spin and its  next nearest
neighbor is plotted as a function of the frustration parameter
$\alpha$ for chain lengths $N=16, 20, 24$ and DM vector $D=0.05$.
In the Luttinger liquid and dimer phases, the correlation between
different components, $g_{lm}^{xy}$, $g_{lm}^{yx}$ is equal to
zero and only in the chiral phase has non-zero value. As it can be
seen from Fig.~\ref{Concurrence}, in the Luttinger liquid
region, $\alpha<\alpha_{c_1}$, since the correlations
$g_{lm}^{xx}$, $g_{lm}^{yy}$ are equal, therefore the concurrence
is equal to zero. On the other hand in the intermediate chiral
region, $\alpha_{c_1}<\alpha<\alpha_{c_2}$, due to non-zero
value of the correlations between different components the
concurrence increases by increasing the frustration and at the
second critical value $\alpha=\alpha_{c_2}$ is maximum. In sector of the dimer ground state, $\alpha>\alpha_{c_2}$,
the correlations $g_{lm}^{xx}$, $g_{lm}^{yy}$ are equal and close
to the maximum value $\frac{1}{4}$\cite{Mahdavifar08-1}. In this
case, as it is clearly seen from the Fig.~\ref{Concurrence}, the
concurrence has the almost constant value that obeys from the
equation $C_{lm}=\mid g_{lm}^{xx}+g_{lm}^{yy} \mid \simeq 0.5$.
The deviation from the saturation value $0.5$ is a result of the
induced quantum fluctuations by DM interaction. In the inset of
Fig.~\ref{Concurrence}, the $N$-dependence of the concurrence is
investigated for different values of the frustration in the dimer
sector of the ground state phase diagram. Finite linear
extrapolated results show that in the thermodynamic limit
$N\longrightarrow \infty$, the next nearest neighbors are really
entangled in the dimer phase. Thus, in the 1D frustrated
ferromagnetic spin-1/2 model in presence of the DM interaction,
only the next nearest neighbors are entangled in the gapped
chiral and dimer phases.

It is expected that quantum nature of entanglement
might provide explanatory and predictive power for the investigation of quantum
phase transitions\cite{Amico08,Osterloh00}.  Thus concurrence -pairwise entanglement- which refers to
quantum correlations has emerged as one of the important
tools. In our model Eq.(\ref{J1-J2-DM}),
only the next nearest neighbors quantum correlation lead to
concurrence have different nonzero value in the gapped
chiral and dimer phases and zero value in the Luttinger liquid phase.

%%%%%%%%%%%%%%%%%%%%%%% BEGIN OF THE FIGURE N 3 %%%%%%%%%%%%%%%%%%%%%%%%%%
%%%%%%%%%%%%%%%%%%%%%%%%%%%%%%%%%%%%%%%%%%%%%%%%%%%%%%%%%%%%%%%%%%%%%%%%%%

%%%%%%%%%%%%%%%%%%%%%%%%%%%%%%%%%%%%%%%%%
%%%%%%%%%%%%%%%%%%%%%%%%%%%%%%%%%%%%%%%%%%%%%%%%%%%%%%

%%%%%%%%%%%%%%%%%%%%%%%%%%%%%%%%%%%%%%%%%
\begin{figure}[t]
\centerline{\psfig{file=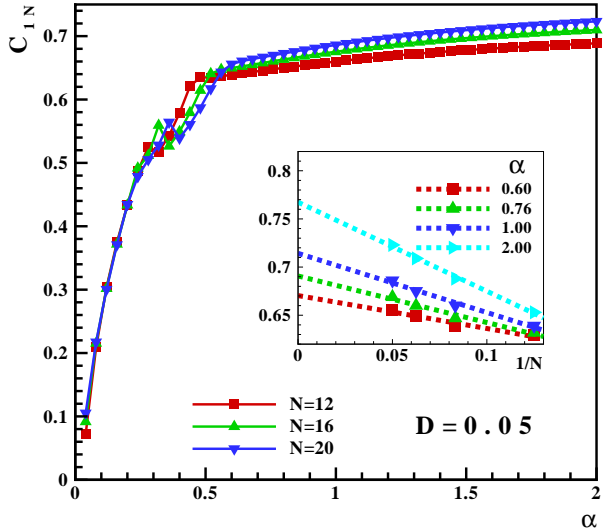,width=3.65in}}
\caption{(color online). The entanglement between end spins as a
function of $\alpha$, with $D=0.05$  for different chain lengths
up to $N=20$, with open boundary condition. The inset shows
$C_{1N}$ as a function of the inverse length $1/N$ for different
values of the frustration $\alpha=0.60, 0.76, 1.00, 2.00$. The
extrapolated results in the thermodynamic limit $N\longrightarrow
\infty$ are less than one.}\label{C1-N}
\end{figure}
%%%%%%%%%%%%%%%%%%%%%%%%%%%%%%%%%%%%%%%%%%%%%%%%%%%%%%

On the other hand, entanglement generation and distribution are problems of central
importance in performing quantum-information(QI) tasks, like
teleportation \cite{DBoschi98}  and quantum
cryptography\cite{NGisin02}. From the QI perspective, it would be
attracting to create sizable entanglement between particles that
are located at a distance larger than a few sites. This fact
naturally leads to the concepts of long distance entanglement
(LDE) as a sort of quantum order parameter\cite{LCampos06}. There
are some models same as the dimerized-frustrated model, spin-1
Heisenberg chain with biquadratic interaction and so on, which
able to produce LDE\cite{LCampos06}. Here we are interested to
check this feature in our model. As we showed in previous sections
our spin frustration model leads to dimerized phase if the
parameter $\alpha$ exceeds a certain critical value
($\alpha_{c_2})$,  so the process of dimerization with the open
boundary condition enables frustrated model to produce LDE. In
Fig.~\ref{C1-N}, we give the numerical Lanczos results on the long
distance entanglement of frustrated Heisenberg chain in the
presence of the DM interaction, particularly $D=0.05$. As it can
be seen from Fig.~\ref{C1-N}, in contrast next-nearest neighbor
entanglement, the system has got nonzero LDE which increases
rapidly by increasing $\alpha$ up to $\alpha_{c_2}$
and then after that it reaches its saturation value about
$C_{1-N}\simeq0.7$. In the inset of Fig. \ref{C1-N}, the
$N$-dependence of the LDE is investigated for different values of
the frustration in the dimer sector of the ground state phase
diagram. Linear extrapolated results show that in the
thermodynamic limit $N\longrightarrow \infty$, the end-spins are
 entangled in the dimer phase for open boundary conditions.

\section{Conclusion}\label{sec VI }

To summarize, we studied the effect  of a uniform
Dzyaloshinskii-Moriya (DM) interaction on the ground state phase
diagram of the one-dimensional (1D) isotropic frustrated
ferromagnetic spin-1/2 model using the analytical cluster method
and numerical Lanczos technique. Our classical analysis results based
on LK method show
that immediately after turning on the DM interaction, the ground
state of the system goes to the chiral phase, independent of the
frustration parameter $\alpha=\frac{J_{2}}{|J_{1}|}$. Thus, we
concluded that the classical ground state phase diagram of the 1D
frustrated ferromagnetic spin-1/2 model with added uniform DM
interaction consists of a single phase: "chiral".

To find the quantum corrections of adding the DM interaction, we did a
very accurate numerical experiment.  We implemented the Lanczos
algorithm to find the ground state in finite chains. Based on the
 numerical results of the order parameters and correlation functions, we
 identified three different phases. In the region $\alpha<\alpha_{c_1}$,
 the ground state of the system is in the Luttinger liquid phase.
 In the region $\alpha>\alpha_{c_2}$ the true long-range dimer order
 exists in the ground state phase diagram and in the
 intermediate region $\alpha_{c_1}<\alpha<\alpha_{c_2}$, the
 system finds in the gapped chiral phase.

 On the other hand, we tried to answer this question: "Are the spins entangled in different
 sectors of the ground state phase diagram?"  Our numerical results
 for the entanglement showed that the nearest neighbor spins are not
 entangled in all regions of the quantum ground state phase diagram. But, the next nearest neighbor
 spins are entangled in the gapped chiral and dimer phases, and became saturate in
 the region $\alpha>\alpha_{c_2}$,  which can be take it as another witness the presence of dimer phase.
 We have even checked this model with open boundary conditions as an candidate for LDE model,
 where calculations showed that  the model reaches its saturated value around  $\alpha_{c_2}$ which is
 more than next nearest neighbor entanglement case $C_{1N}>C_{j~j+2}$.

 %%%%%%%%%%%%%%%%%%%%%%%%%%%%%%%%%%%%%%%%%%%%%%%%%%%%%%%%%%%%%%%%%%%%%%%%%%%%%%%%%%%%%%%%%%
\section{Acknowledgments}
It is our pleasure to thank T. Kaplan, G. I. Japaridze, T. Vekua,
A. Akbari, J. Abouie, S. H. Sadat, M. R. Soltani, G. Rigolin, T. Nishino,  and T.
Hikihara  for very useful comments and useful suggestions. We are also grateful
to S. N. Rasouli for reading our manuscript.
%%%%%%%%%%%%%%%%%%%%%%%%%%%%%%%%%%%%%%%%%%%%%%%%%%

%%%%%%%%%%%%%%%%%%%%%%%%%%%%%%%%%

%%%%%%%%%%%%%%%%%%

%-----------------------------------------------------------------------------
\vspace{0.3cm}
%\section*{References}

\end{document}